\documentclass[twocolumn, prl, superscriptaddress]{revtex4}

\usepackage{amsmath}
\usepackage{subfigure}
\usepackage{natbib}

\newcommand{\ms}[1]{{\color{blue} #1}}

\newcommand{\fixme}[1]{{\color{red}[Fixme: #1]}}

\usepackage{graphicx}

\usepackage[a4paper, total={7in, 10in}]{geometry}

\usepackage{hyperref}
\hypersetup{
    colorlinks=true,
    linkcolor=blue,
    filecolor=magenta,      
    urlcolor=blue,
    pdftitle={Overleaf Example},
    pdfpagemode=FullScreen,
    }

\usepackage{dcolumn}
\usepackage{bm}

\usepackage{orcidlink}

\begin{document}

\title{Discovering the Cosmic Evolution of Supermassive Black Holes using Nano-Hertz Gravitational Waves and Galaxy Surveys}

\author{Mohit Raj Sah \orcidlink{0009-0005-9881-1788}}\email{mohit.sah@tifr.res.in}
\author{Suvodip Mukherjee \orcidlink{0000-0002-3373-5236}}\email{suvodip@tifr.res.in}
\affiliation{Department of Astronomy and Astrophysics, Tata Institute of Fundamental Research, Mumbai 400005, India}

\begin{abstract}
The formation and evolution of supermassive black holes (SMBHs) in the Universe remains an open question in cosmology. We show for the first time that the evolution of SMBHs with redshift leads to a unique signature on the angular cross-correlation power spectrum between the multi-frequency nano-hertz (nHz) stochastic gravitational wave (SGWB) and the galaxy density in the Universe. By using galaxy catalogs from the upcoming Rubin LSST Observatory in synergy with the nHz SGWB signal accessible from the Square Kilometer Array, we can measure this signal with a signal-to-noise ratio above five, thereby opening a new observational window to the cosmic evolution of SMBHs across redshift. This discovery space that can be opened by the cross-correlation of the nHz SGWB will not be possible by any other currently known techniques.

\end{abstract}

\maketitle
\textbf{\textit{Introduction: }}
The Pulsar Timing Array (PTA) consortia, which include the North American Observatory for Gravitational Waves (NANOGrav; \cite{mclaughlin2013north}), the European PTA (EPTA; \cite{desvignes2016high}), the Parkes PTA (PPTA; \cite{manchester2013parkes}), the Indian PTA (InPTA; \cite{joshi2018precision}), and the Chinese PTA (CPTA; \cite{xu2023searching}), have recently announced evidence of a stochastic gravitational wave background (SGWB) in the $10^{-9}-10^{-7}$ Hz range \citep{agazie2023nanograv,antoniadis2023second,zic2023parkes,xu2023searching}. The most promising source of the SGWB in the PTA band is the population of the supermassive black hole binaries (SMBHBs). Other potential sources of the nano-hertz (nHz) SGWB are the GWs generated by physical phenomena in the early Universe, like cosmic strings, phase transitions, domain walls, etc \citep{ellis2024source,madge2023primordial}. 

These SMBHBs form following the merger of two galaxies, each hosting a supermassive black hole (SMBH). These binary systems dissipate energy through various mechanisms, including dynamical friction, stellar loss-cone scattering, and viscous drag before their separation is small enough to emit GW radiation in the PTA band \citep{sampson2015constraining,kelley2017massive,chen2017efficient,izquierdo2022massive}. The superposition of GWs from all such SMBHBs that are unresolved by the detector generates the SGWB. Since only a fraction of galaxies are expected to host SMBHBs with separations close enough to emit in the PTA band, the SGWB is likely to exhibit spatial anisotropy \citep{sato2023exploring,gardiner2024beyond,sah2024imprints}.  Moreover, the angular power spectrum ($C_{\ell}^{\rm GWGW}$) of the SGWB, is likely to be dominated by shot noise due to fewer number of sources, hence, the $C_{\ell}^{\rm GWGW}$ spectrum is expected to be constant for $\ell$ $>$ 0 \cite{sah2024imprints}. As a result, the information on the correlation of GW sources with galaxies cannot be explored using $C_{\ell}^{\rm GWGW}$. However, the formation of the SMBHBs and their connection with galaxies is imprinted in the nHz GW anisotropy signal. Measuring it will provide a wealth of information that can help us understand the longstanding cosmological problem of the formation and evolution of SMBHs. \cite{sah2024imprints}. 

\begin{figure}
    \centering    \includegraphics[width=9cm]{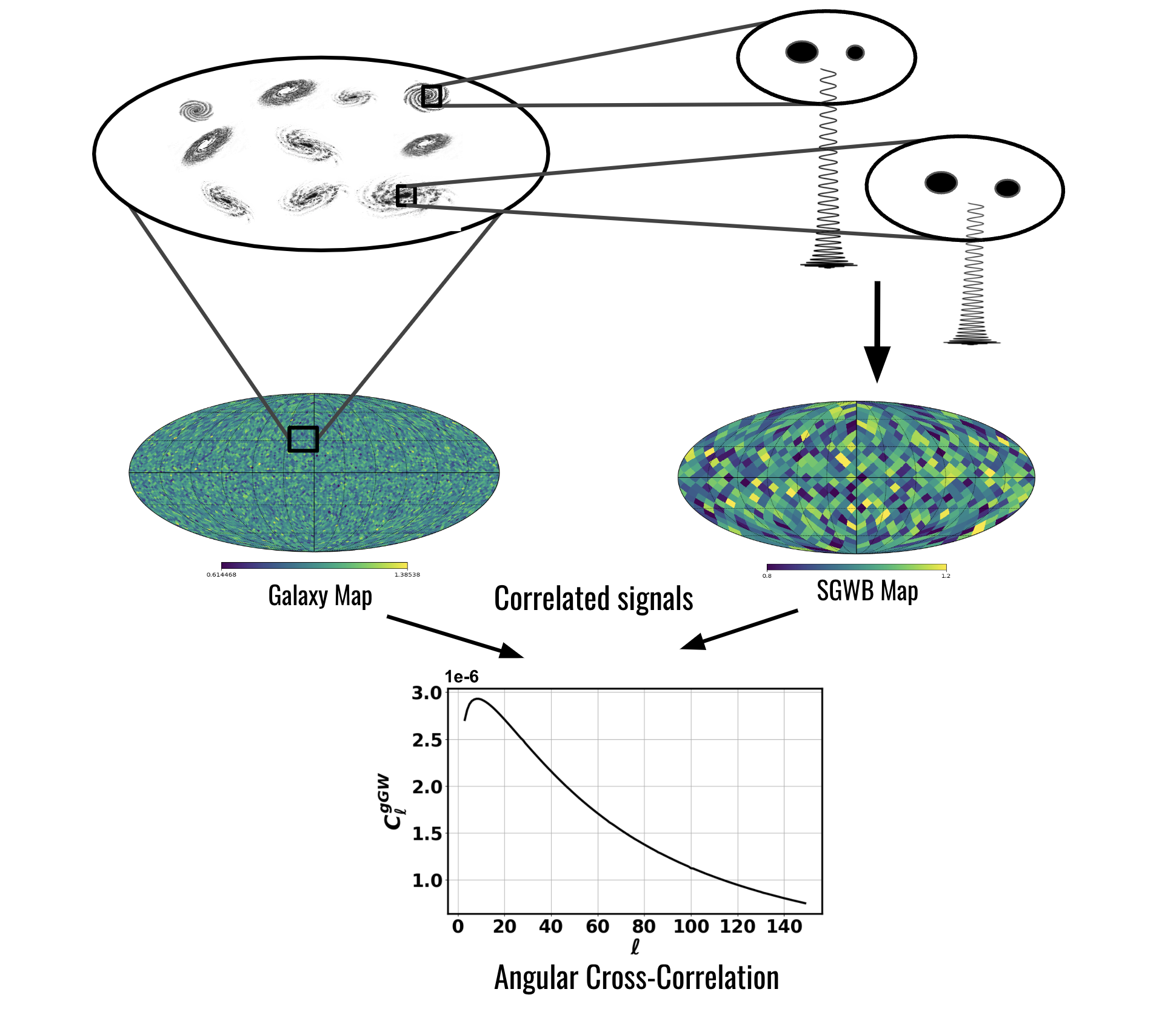}
    \caption{A schematic diagram depicting the origin of correlation between the SGWB and the galaxy distribution. The SMBHBs reside at the center of galaxies and are spatially correlated. As a result, the co-evolution of the SMBHs and galaxies with redshift will exhibit a non-zero spatial angular cross-correlation power spectrum between SGWB and galaxies for different astrophysical scenarios.}
    \label{Mot1}
\end{figure}

We propose a new way to infer this signal from the nHz GWs using cross-correlation with galaxy distribution, which is not dominated by the shot noise.  In Fig. \ref{Mot1} we illustrate the idea of the cross-correlation. Since the SMBHBs reside at the center of the galaxies, the spatial distribution of the SGWB is likely to follow the distribution of the galaxies. Due to the finite number of the SMBHBs contributing to the SGWB signal, the clustering anisotropic signal of SGWB is going to be dominated by the shot noise. However, by cross-correlating the SGWB anisotropy with the galaxy catalog we can enhance the clustering signal of the SGWB.

Cross-correlating the SGWB data with large-scale structure data from galaxy surveys will not only allow us to detect the anisotropy but also provide valuable insights into the formation and evolution of SMBHBs and their host galaxies. The redshift evolution of the population of the SMBHBs will have a significant imprint on the cross-correlation signal. This can help us understand the merger history of SMBHBs over cosmic time. Therefore, the cross-correlation signal will complement the information from the auto-correlation signal ($C_{\ell}^{\rm GWGW}$), as discussed in \cite{sah2024imprints}.

\textbf{\textit{SGWB from SMBHBs population:}} Most of the SMBHs that are expected to contribute to the PTA observations will appear practically monochromatic within our observation period. This is because the orbits of most of the binaries are expected to be very stable. The superposition of GW signals from these unresolved monochromatic nHz GW sources gives rise to the SGWB signal. Mathematically, the SGWB density is defined as the GW energy density per unit logarithmic frequency divided by the critical energy density of the Universe. The SGWB density due to the sources in unit solid angle can be written as \citep{phinney2001practical,sah2024imprints}

\begin{equation}
    \begin{aligned}
     \Omega_{\rm gw}(f, \hat n) = \frac{1}{\rho_{\rm c} c^2} \int d\tilde{V} & \int \prod_{i}^{n}  d\theta_i \Big[\kappa(z,\Theta_{\rm n},f_{\rm r})  n_{\rm v}(z,\hat{n})\Big]\\
     &\times \Big[ \frac{1}{4 \pi d_{\ell}^{2}(z) c} \frac{ dE_{\rm{gw}}(f,\Theta_{\rm n}, \hat{n})}{{ dt_{\rm r}}}\Big],
    \end{aligned}
    \label{SGWB2}
\end{equation}
where $f_{r}$ represents the source frame frequency, $d_{\ell}(z)$ represents the luminosity distance of the source, $\frac{d\tilde{V}}{dd_{\ell}} = \frac{d^2_{\ell}(z)}{(1+z)^3} $,   $\kappa(z,\Theta_{\rm n},f_{\rm r})$ represents the mean number of the SMBHBs hosted by each galaxy with source properties $\Theta_{\rm n}$, emitting GW in unit logarithmic frequency bin, at redshift z, $n_{\rm v}(z,\hat{n})$ represents the number of galaxies per unit comoving volume at redshift z in the direction $\hat n$, $\frac{ dE_{\rm{gw}}(f,\Theta_n,\hat n)}{{ dt_{\rm r}}}$ is the GW energy flux emitted by the source per unit time, and $\Theta_{\rm n}$ = $\{\theta_i\}_{i=1}^{n}$ denotes the GW source parameters. The relative fluctuation in the frequency integrated $\Omega_{\rm gw}(f,\hat{n})$ can be defined as

\begin{equation}
    \begin{aligned}
    \Delta\bm{\Omega}_{\rm GW}(\hat{n}) \equiv  & {\frac{ \int \frac{df}{f} ~  \Omega_{\rm gw}(f,\hat{n}) - \int \frac{df}{f}  ~ \overline{\Omega}_{\rm gw}(f)}{\int \frac{df}{f} ~\overline{\Omega}_{\rm gw}(f)}}\\    
    = &\int ~dz~ 
    \delta_{\rm m}(z,\hat{n}) ~ b_{\rm g}(z) ~  \phi_{\rm GW}(z),
    \end{aligned}
    \label{SGWB_Fluc}
\end{equation}
here $\overline{\Omega}_{\rm gw}(f)$ is the mean SGWB density per unit solid angle, $\delta_{\rm m}(z,\hat{n})$ is the matter density fluctuation given by, $\delta_{\rm m}(z,\hat{n}) = \frac{n_{\rm m}(z,\hat{n})}{\bar{n}_{\rm m}(z)} -1 $, where $n_{\rm m}(z,\hat{n})$ is the matter density at redshift z in the direction $\hat{n}$, and  $\bar{n}_{\rm m}(z)$ is the mean matter density at redshift z. $b_{\rm g}(z)$ is  the galaxy bias, and $\phi_{\rm GW}(z)$ is given by

\begin{equation}
\phi_{\rm GW}(z)  = \frac{\int  \prod\limits_{i}^{n} \frac{d\theta_i}{ d_{\ell}^{2} } \int \frac{df}{f} ~ \Big[\kappa(z,\Theta_{\rm n},f_{\rm r})  \frac{dn}{dz}\Big] \Big[\frac{ dE_{\rm gw}(f,\Theta_{\rm n}, \hat n)}{dt_{\rm r}}\Big]}{\int \frac{dz}{d_{\ell}^{2}}\int  \prod\limits_{i}^{n} d\theta_i \int \frac{df}{f} ~  \Big[\kappa(z,\Theta_{\rm n},f_{\rm r})  \frac{dn}{dz}\Big] \Big[\frac{ dE_{\rm gw}(f,\Theta_{\rm n}, \hat n)}{{ dt_{\rm r}}}\Big]},
\label{phiGW}
\end{equation}

where $\frac{dn}{dz}$ is number of galaxies per unit redshift bin. Instead of a frequency-integrated anisotropy map $\Delta\bm{\Omega}_{\rm GW}(\hat{n})$, one can also use an anisotropic map at every frequency if the noise is low in the individual map in comparison to the signal. In this analysis, we explore a frequency-integrated map. However, a cross-correlation analysis as a function of frequency can be trivially extended.

Simulations and observational studies provide strong evidence for a correlation between the mass of SMBHs and the properties of their host galaxies \citep{haring2004black,jahnke2011non,mcconnell2013revisiting,reines2015relations,izquierdo2022massive,hoshi2024relationship}. A previous study based on Romulus simulation \citep{tremmel2017romulus,tremmel2020formation,saeedzadeh2023cool} has demonstrated that the majority of SMBHBs relevant to the PTA band are located in galaxies with high stellar mass and low star formation rates. SMBHBs with masses greater than $10^{8} M_{\odot}$ are more frequently found in galaxies with stellar masses exceeding $10^{11} M_{\odot}$ \cite{saeedzadeh2023shining}.

The population of the SMBHBs appearing in Eq. \eqref{SGWB2} can then be written as \citep{sah2024imprints} 
\begin{equation}
    \begin{aligned}
        \kappa(z,\Theta_{\rm n},f_{\rm r}) \propto   ~  & P( M_{\rm BH}| M_*,z) P(q| M_*,z) P(f_r| M_*,z),
    \end{aligned}
    \label{pop}
\end{equation}
where $M_{\rm BH}$ is the primary mass, q is the mass ratio of the SMBHB, and $M_{*}$ is the host galaxy stellar mass. We model $P( M_{\rm BH}| M_*,z)$ as
\begin{equation}
    P( M_{\rm BH}|  M_{*},z) =  \mathcal{N}( \mathrm{Log}_{10}(M_{\rm BH})| \mathrm{Log}_{10}(M_{\rm \mu}),\sigma_{\rm m}),
    \label{M1}
\end{equation}
where $\mathcal{N}$ denotes a normalised Gaussian with standard deviation  $\sigma_{\rm m}$ and mean $ \mathrm{Log}_{10}(M_{\rm \mu})$  given by
\begin{equation}
\mathrm{Log}_{10}(M_{\rm \mu})= \eta + \rho ~ \mathrm{Log}_{10}( M_{*}/10^{11} M_{\odot}) + \nu  z,
\label{Mmu}
\end{equation}
where $\eta$, $\rho$, and $\nu$ are free parameters, where $\nu$ controls the redshift evolution of the relation \cite{sah2024imprints}. The probability distributions for the mass ratio, $P(q | M_, z)$, and $P(f_r | M_, z)$, are modeled as
\begin{equation}
     P(q|  M_{*},z) \propto \bigg\{
    \begin{array}{cl}
    & 1/q, \quad  0.01 < q < 1,\\
    & 0, ~~ \rm{else}, 
    \end{array}
    \label{q}
\end{equation}
\begin{equation}
    P( f_r|  M_{*},z) \propto  \frac{dN_{\rm f}}{dt_{\rm r}} \frac{dt_{\rm r}}{df_{\rm r}}, 
    \label{freq_dist}
\end{equation}
where $\frac{dt_{\rm r}}{df_{\rm r}}$ = $f^{-11/3}_{\rm r}$ is the GW residence time of the binary at rest frame frequency $f_{\rm r}$, and $\frac{dN_{\rm f}}{dt_{\rm r}}$ is the coalescence rate of the binary emitting at frequency $f_{r}$. We model $\frac{dN_{\rm f}}{dt_{\rm r}}$ as
\begin{equation}
    \frac{dN_{\rm f}}{dt_{\rm r}} \propto f_{\rm r}^{\delta_{\rm f} (z-z_0)^{\mu}},
\end{equation}
where the parameters $\delta_{\rm f}$, $z_0$, and $\mu$ control the frequency and the redshift dependence of the coalescing rate across cosmic time.

\textbf{\textit{Cross-correlation formalism:}} The study of the anisotropy in the SGWB is crucial for understanding the nature and the origin of the SGWB signal and its connection with the evolution of the galaxies. We propose to investigate the anisotropy in the SGWB using galaxy catalogs by measuring the angular cross-correlation power spectrum. The SGWB fluctuation ($\Delta\bm{\Omega}_{\rm GW}(\hat{n})$) and the galaxy density fluctuation ($\delta_{\rm g}(\hat{n})$) can be written in the spherical harmonic basis as  
\begin{align}
    \Delta\bm{\Omega}_{\rm gw}(\hat{n}) =& \sum\limits_{\ell} \sum\limits_{m = -\ell}^{\rm \ell}  a_{\ell m}^{\rm GW} Y_{\ell m}(\hat{n}),\\\nonumber
    \delta_{\rm g}(\hat{n}) =& \sum\limits_{\ell} \sum\limits_{m = -\ell}^{\rm \ell} a_{\ell m}^{\rm g} Y_{\ell m}(\hat{n}),
\end{align}
where $a^{X}_{lm}$ (X = g,GW) is given by 
\begin{equation}
    a^{X}_{lm} = 4 \pi i^{\ell} \int dz \phi_{\rm X}(z) \int \frac{d^3k}{(2\pi)^{3}}  ~ \delta_{\rm X}(k,z) ~ j_{\ell}(k r(z)) ~ Y^{*}_{\ell m}(\hat{k}). 
\end{equation}
Using this equation, we can write the theoretical angular cross-correlation power spectrum between the nHz GW background signal and the galaxy distribution, $C_{\ell}^{\rm gGW}$, as follows

\begin{equation}
    \begin{aligned}
       C_{\ell}^{\rm gGW}\equiv\, &
       \langle a_{\ell m}^{\rm GW}  a_{\ell 
        m}^{*\rm g}\rangle,\\
         =& \frac{2}{\pi}  \int dz_1 W_{\rm g}(z_1) D(z_1) \int dz_2 W_{\rm GW}(z_2) D(z_2)  \\
         & \times ~ \int k^2 dk ~  j_{\ell}(kr(z_1)) j_{\ell}(k r(z_2))  ~ P(k),
    \end{aligned}
    \label{ClgGW_1}
\end{equation}
where, $j_{\ell}(kr)$ is the spherical Bessel functions of order $\ell$, and $\phi_{\rm g}(z)$ is the galaxy radial selection function given by
\begin{equation}
    \phi_{\rm g}(z) \equiv \frac{\frac{dn}{dz}}{\int \frac{dn}{dz} ~ dz},
\label{phig}
\end{equation}
where $W_{\rm g}(z) = \phi_{\rm g}(z)  b_{\rm g}(z)$ and $W_{\rm GW}(z) = \phi_{\rm GW}(z) b_{\rm g}(z)$ are the galaxy and GW window functions respectively,  $P(k)$ is the matter power spectrum at z = 0, and D(z) is the factor that describes the growth of density perturbations in the universe over time. It is defined as the ratio of the amplitude of density perturbations at a given redshift z to the amplitude of those perturbations at a reference redshift. It can be calculated by solving the linear perturbation theory equations for the growth of matter density perturbations \citep{dodelson2020modern,peebles1980large}.

In the near future, PTAs will primarily be able to measure the signal at large angular scales. Therefore, in this paper, we will focus on the low $\ell$ part. We will estimate the signal at low $\ell$ using the theoretical technique. We show the validation of the theoretical estimate by applying it on the \texttt{MICECAT} simulations \citep{crocce2015mice} in the appendix using the multi-scale adaptive technique \cite{sah2024imprints}.

\begin{figure}
    \centering
    \includegraphics[width=8cm]{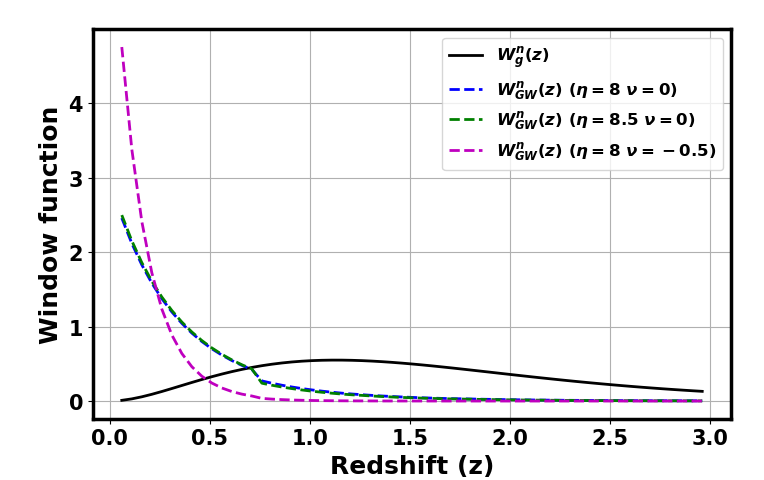}
    \caption{Normalised window function for LSST like galaxy catalog, $W^{\rm n}_{\rm g}$, represented by a solid line, and frequency integrated window function for SGWB, $W^{\rm n}_{\rm GW}$, represented by dash lines for different $M_{*}-M_{\rm BH}$ relations with fixed $\rho = 1$. The window functions for the cases $\eta = 8$, with $\nu =0$ and $\eta = 8.5$ with $\nu =0$ are overlapping.}
    \label{Wn2}
\end{figure}

\begin{figure}
    \centering
    \includegraphics[width=8cm]{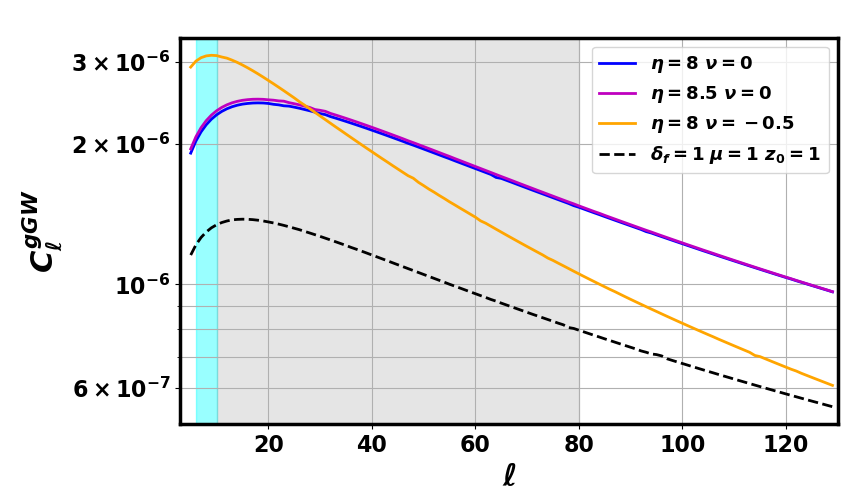}
    \caption{Theoretical $C_{\ell}^{\rm gGW}$ for LSST-like surveys for different population models. The solid lines represent the case with different $M_{*}-M_{\rm BH}$ relation. The dash line represents the case with redshift-dependent frequency distribution, i.e., $\mu = 1$. The cyan-shaded region represents the $\ell$ modes accessible by the current PTA and the grey-shaded region represents the $\ell$ modes accessible with the integration of the SKA into PTA.}
    \label{ClgGW_L}
\end{figure}

\textbf{\textit{ Expected cross-correlation signal between nHz GWs and LSST-like galaxy surveys:}}
In this section, we evaluate the potential for measuring the angular cross-correlation signal $C_{\ell}^{\rm gGW}$ 
 between the SGWB and galaxy distributions using anticipated number density (dn/dz) for future Rubin Large Synoptic Survey Telescope (LSST)-like Observatory \citep{ivezic2019lsst}. The LSST, with its exceptional depth and wide field of view, is expected to cover approximately 18,000 square degrees of the sky and provide a highly detailed map of the galaxy distribution up to a redshift of $z\sim3$, significantly enhancing our ability to detect and analyze such cross-correlation signals. Other notable future surveys include the Dark Energy Spectroscopic Instrument (DESI) \citep{dey2019overview}, the Euclid mission \citep{collaboration2022euclid}, and the Spectro-Photometer for the History of the Universe, Epoch of Reionization, and Ices Explorer (SPHEREx) \citep{korngut2018spherex}, etc. Together, these upcoming surveys will enable a more accurate and detailed analysis of the cross-correlation signal. We choose to focus on LSST over other future surveys because of its ability to cover very high redshifts and large sky areas, making it particularly well suited for studying the cross-correlation with nHz across a broad range of scales and redshift bins.

To estimate $C_{\ell}^{\rm gGW}$ for LSST-like surveys, we use the redshift distribution of the source, $\frac{dn}{dz}$, in Eqs \eqref{phiGW} and \eqref{phig} to estimate window functions, $W_{\rm gGW}(z)$ and $W_{\rm g}(z)$ from the simulated LSST catalog. To estimate the SMBH mass, we need the stellar mass of the host galaxy. We determine the stellar mass function using a redshift-dependent Schechter function, as described by \cite{mcleod2021evolution}. This function accounts for the evolution of the stellar mass distribution with redshift, providing a realistic representation of the galaxy population over redshift. Using the relation given by Eqs \eqref{M1} and \eqref{Mmu}, we then estimate the mass distribution of the source SMBHBs from the galaxy mass stellar function. This relation connects the stellar mass of a galaxy to the mass of its central SMBH, allowing us to model the SMBHB population.

In Fig. \ref{Wn2}, we show the normalized window function of the galaxy ($W^{\rm n}_{\rm g}$) as well as the SGWB ($W^{\rm n}_{\rm GW}$) for LSST like surveys. The window functions offer insights into the distribution and relative contributions of these sources across different redshifts. The window function of the galaxy peaks around z = 1 and then starts to fall above z = 1. However, the SGWB window function exhibits a profile that gradually decreases with increasing redshift. This trend arises because sources at lower redshifts appear brighter than those at higher redshifts. In addition, the delay time between the merger of the galaxy and the merger of the SMBHBs is of the order of a fraction of the age of the Universe \citep{saeedzadeh2023shining}, leading to an expectation of more SMBHBs being found at low redshifts. In the scenario with lighter SMBHs at higher redshifts ($\nu=-0.5$), the $W^{\rm n}_{\rm GW}$ function has a much higher value at low redshifts compared to higher redshifts than in other models. This is because, in these cases, the GW signal from high redshift sources is fainter than the redshift independent cases ($\nu=0$). Other scenarios of SMBH evolution are also possible and can be explored by this technique.

 In Fig. \ref{ClgGW_L}, we show the theoretical prediction for the cross-angular power spectrum $C_{\ell}^{\rm gGW}$ 
based on the anticipated number density of galaxies from LSST-like surveys. The theoretical curve demonstrates an expected correlation between the SGWB and the galaxy distribution at various angular scales. We illustrate the $C_{\ell}^{\rm gGW}$  curve for different population models. The solid lines represent the scenario with different $M_{*}-M_{\rm BH}$ relations. In this case, we vary the parameters $\eta$ and $\nu$  while keeping the parameters  $\rho$ = 1, $\delta_f$ = 1 and $\mu=0$ fixed. In contrast, the dash lines illustrate the scenario with a redshift-dependent frequency distribution, i.e., $\mu=1$, with $\eta$ = 8, $\nu$ = 0, $\rho$=1, $\delta_f$ = 1, and $z_0$ = 1. Additionally, we show the $\ell$ mode accessible by the current PTA by cyan-shaded region and accessible with the integration of SKA in the PTA by grey-shaded region. In the SKA era, we will be able to time around 6000-millisecond pulsars\citep{smits2009pulsar}, making it possible to go up to high $\ell$-modes.

The curve remains consistent despite changes in the $\eta$ parameter. The $\eta$ parameter represents the minimum mass threshold of galaxies required to host the SMBHBs that are significant for PTA observations. The cross-correlation is not affected significantly by the variation in the $M_{*}-M_{\rm BH}$ relation for the redshift-independent cases. However, the cross-correlation changes significantly when redshift evolution is introduced into the $M_{*}-M_{\rm BH}$ relation. With $\nu = -0.5$, we have heavier mass SMBHs at lower redshifts, substantially increasing the contribution of low-redshift sources compared to high-redshift sources. This increased contribution from lower redshifts corresponds to regions of higher clustering, as the clustering of the universe decreases with the redshift, resulting in a higher cross-correlation signal. This implies the cross-correlation signal will hold the unique signature of the redshift evolution of the SMBH population in the Universe on the angular cross-correlation power spectrum $C_{\ell}^{\rm gGW}$. Therefore, by analyzing the signature, we can gain a deeper understanding of the SMBHB population and its evolution. Additionally, the case where we have included the redshift-evolution of the frequency distribution has a lower cross-correlation signal.
This is because the coalescence rate for this case decreases with source frame frequency for $z<1$, making the frequency distribution steeper towards high frequencies, especially at low redshifts. This means fewer sources are present at low redshift hence decreasing the contribution. It is important to point out that, the cross-correlation signal explores additional information on the SMBHs evolution which cannot be measured from the auto-correlation as it will be dominated by the shot-noise term \cite{sah2024imprints}.

\textbf{\textit{Discovery-space in synergy between SKA and LSST: }}
The signal-to-noise ratio (SNR) of the $C_{\ell}^{\rm gGW}$ by combining signals from individual GW frequency bins can be written as 
\begin{equation}
     SNR = \sqrt{\sum\limits_{\ell} \Big[\frac{(C_{\ell}^{\rm g,\rm GW})^{2}}{\Sigma_{gGW}^2(\ell)}\Big]},
     \label{SNR_eq}
\end{equation}
where $ \Sigma_{\rm GW}(\ell)$ is the uncertainty in the measurement of the $C_{\ell}^{\rm gGW}$, which is given by 
\begin{equation}
    \begin{aligned}
        \Sigma_{\rm gGW}^{2}(\ell) =  &\frac{1}{f_{\rm sky}   (2 \ell+1)}  \bigg[(C_{\ell}^{\rm gGW})^2 \\ &  + (C_{\ell}^{\rm gg} + \frac{1}{n_{\rm g}})  (C_{\ell}^{\rm GWGW} + N_{\ell})\bigg], 
    \end{aligned}
    \label{Var}
\end{equation}
where $C_{\ell}^{\rm gg}$ and $C_{\ell}^{\rm GWGW}$ are the galaxy and GW angular auto-correlation power spectrum respectively, $n_{\rm g}$ is the galaxy mean number density (per steradian), $f_{\rm sky}$ is the sky fraction covered by the galaxy catalog. As we want to discover the anisotropy in the nHz, under the hypothesis of isotropic SGWB maps of the simulations, we take $C_{\ell}^{\rm GWGW}(\ell>0)=0$ and $C_{\ell}^{\rm gGW}=0$. The term $N_{\ell}$ is the noise power spectrum in the measurement of SGWB anisotropy due to pulsar noise, instrument noise, and also the amplitude of background SGWB \cite{siemens2013stochastic} as
\begin{equation}
     N_{\ell} = \sigma^2 e^{\frac{\ell (\ell+1) \theta^2_{\rm b}}{8\rm ln2}},
\end{equation}
where $e^{\frac{\ell (\ell+1) \theta^2_{\rm b}}{8\rm ln2}}$ is the Gaussian beam window function that takes in account the resolution of the instrument, $\theta_{\rm b}$ is the full-width half maxima of the window function, and $\sigma^2$ represents the variance in the measurement of the $\Delta\bm{\Omega}_{\rm GW}(\hat{n})$ in units of steradian. The derivation of $N_{\ell}$ is given in the appendix.

The value of $\sigma$ is influenced by several factors, including the number of pulsars, their distribution in the sky, the levels of both white and red noises \citep{NANOGrav:2023ctt,alam2020nanograv}, and total observation time (T). Increasing the number of pulsars and reducing the instrumental noise can decrease the magnitude of $\sigma$ if the pulsars are distributed homogeneously. If the pulsar distribution is inhomogeneous, the pixel noise will be anisotropic. In our analysis, we have chosen $\sigma$ and $l_{\rm max}$ as the free parameters that directly impact the cross-correlation analysis. For both homogeneous and inhomogeneous distributions of pulsars, their timing-residual noise (white and red), and their numbers, the value of $\sigma$ and $l_{\max}$ will vary. For an ideal situation of all pulsars having the same timing residual noise and distributed isotropically in the sky, $\sigma^2= 4\pi \sigma^2_{\rm pix}/N_{\rm pix} $ and $\ell_{\rm max}= \sqrt{\text{Number of Pulsars ($N_{\rm p}$)}}$.

\begin{figure}
    \centering
    \includegraphics[width=9.5cm]{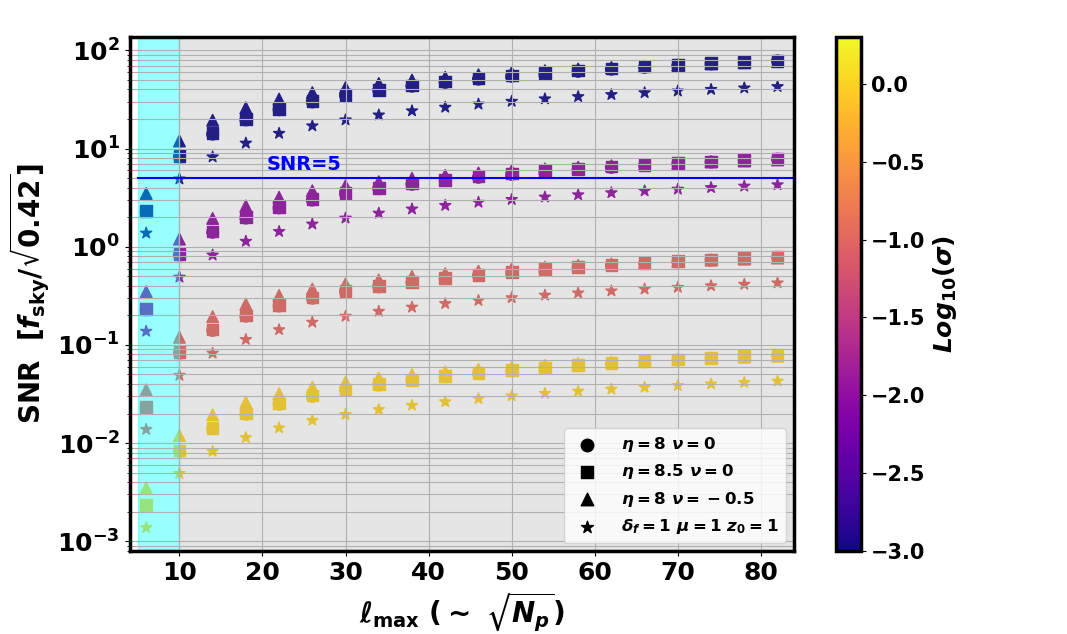}
    \caption{SNR vs $\ell_{\rm max}$ for different $M_{*}-M_{\rm BH}$ relation and pixel uncertainties ($\sigma$, shown in colorbar). The dotted, square, and triangle curves represent the cases where $\eta = 8$ with $\nu=0$, $\eta = 8.5$ with $\nu=0$, and $\eta = 8$ with $\nu=-0.5$ respectively. In these three cases we fix $\rho = 1$, $\delta_f = 1$ and $\mu = 0$. The curves marked by asterisk represent the case with redshift-dependent frequency distribution, i.e., $\mu = 1$, with $\eta$ = 8, $\nu$ = 0, $\delta_f$ = 1, and $z_0$ = 1. The cyan-shaded region represents the maximum $\ell$ modes accessible by the current PTA and the grey-shaded region represents the maximum $\ell$ modes accessible with the integration of the SKA into PTA for the best-case scenario. The blue horizontal line marks the SNR = 5.}
    \label{SNR1}
\end{figure}

In Fig. \ref{SNR1}, we present the SNR measurement for different astrophysical models (shown by different markers) for 4 different values $\sigma$ (varying from 1, 0.1, 0.01, and 0.001), for $f_{\rm sky} =0.42$ (fraction of sky area expected to be covered by LSST survey). The $\ell_{\rm max}$ in the x-axis represents the maximum $\ell$-mode accessible by the detector configuration. The cyan and gray shaded regions represent the maximum $\ell$-mode achievable by the current PTA and SKA respectively. Currently, PTA is timing close to 100 pulsars \cite{verbiest2024status}. This gives the $\ell_{\rm max}$ = 10 for an isotropic distribution, and in reality up to $\ell_{\rm max}= 5$ \cite{pol2022forecasting}. However, in the era of the SKA, we expect to time around 6000-millisecond pulsars\citep{smits2009pulsar} enabling the measurement of the $\ell_{\rm max}$ maximum up to 80 for a homogeneous distribution of pulsars.

The SNR of the signal with the current pulsar configuration and the instrument noise, we have $\sigma \sim 1$ and $\ell_{\rm max}= 5$. This results in a very low SNR (of the order of $10^{-3} - 10^{-2}$) for the astrophysical models of SMBHs we expect\cite{sah2024imprints}. However, with the addition of the SKA pulsars, this number will significantly increase. Even with just around $1600$ new pulsars ($\ell_{\rm max}= 40$) and $\sigma \sim 0.01$, we expect to reach the SNR of around 5. Furthermore, if the instrumental noise is reduced by an order of magnitude (or the number of pulsars increases to about 6000), we can achieve an order of magnitude improvement in the measurement of the cross-correlation signal. This shows that in the near future, using the upcoming surveys, confirming the discovery of the SMBH evolution will be possible. Instead of an LSST-like survey which covers $42\%$ of the sky, if a full sky galaxy catalog can be used by combining different surveys, then the SNR will be improved by a factor of $1.5$.

Among different astrophysical models considered, the cases with no z-evolution of $M_{*}-M_{\rm BH}$ relation ($\nu = 0$), as expected from their $C_{\ell}^{\rm gGW}$ (see Fig. \ref{ClgGW_L}), have comparable SNR, while the case with negative z-evolution ($\nu = -0.5$) shows higher SNR at lower $\ell_{\rm max}$ compared to the case with no z-evolution of $M_{*}-M_{\rm BH}$ relation because the cross-correlation power spectrum is higher at low-$\ell$ for $\nu = -0.5$ than $\nu = 0$. The case with the z-dependent frequency distribution leads to comparatively low SNR measurement by a factor of about $1.3$ than that of the case with the same $M_{*}-M_{\rm BH}$ relation. The SNR of this case is lower due to its lower cross-correlation signal compared to other cases (see Fig. \ref{ClgGW_L}). These results show the potential of this new technique in inferring the cosmic evolution of SMBHs with high SNR in the future.

\textbf{\textit{Conclusion: }}
In this paper, we calculate for the first time the theoretical cross-angular power spectrum $C_{\ell}^{\rm gGW}$ between the SGWB and galaxy distributions. Our analysis demonstrates the feasibility of detecting anisotropies in the SGWB through the synergy of galaxy surveys such as the Rubin LSST Observatory and SKA data. The detection of the anisotropy in the SGWB will be crucial in understanding the astrophysical nature of the SGWB signal. The amplitude and shape of the angular power spectrum will give insight into the growth and evolution of the SMBHB in the Universe. It will show us how fast or slow the growth of the SMBHs is taking place in the Universe. The SNR estimate shows that different astrophysical scenarios of SMBH growth with redshift will have distinct measurability by the cross-correlation studies. 

The estimated SNRs of the $C_{\ell}^{\rm gGW}$ for different levels of pulsar noise and varying numbers of pulsars, in synergy with LSST-like surveys are shown in Fig. \ref{SNR1}. We found that with optimal PTA configurations from SKA and future galaxy survey data, it is feasible to achieve high SNR, thereby making robust measurements of the cross-correlation signal. This will greatly enhance our capability to characterize the evolution of SMBHs with galaxies.

In the future,  we plan to develop a statistical framework that can study the cross-correlation of nHz SGWB with galaxies, which is beyond the current techniques that only explore the auto-correlation of SGWB anisotropy. The application of this technique to the current data will enable to detect or rule out scenarios of extremely large anisotropic SGWB signal. We aim to implement the cross-correlational framework on the  PTA data with the current galaxy catalogs, enabling the first measurement of the cosmic evolution of SMBHs. In summary, exploring the cross-correlation of GW sources with galaxies will open a discovery space on high redshift Universe and SMBHs evolution.  Furthermore, this cross-correlation signal brings a discovery space that cannot be explored from the auto-correlation of the anisotropic maps in the nHz SGWB, as it is impacted by the shot noise of the nHz signal.

\section*{Acknowledgments}
This work is a part of the $\langle \texttt{data|theory}\rangle$ \texttt{Universe-Lab} which is supported by the TIFR and the Department of Atomic Energy, Government of India. The authors would like to thank the $\langle \texttt{data|theory}\rangle$ \texttt{Universe-Lab} for providing computing resources. The authors would also like to acknowledge the use of the CosmoHub \citep{carretero2017cosmohub,tallada2020cosmohub} for galaxy catalog as well as the following Python packages in this work: Numpy \citep{van2011numpy}, Scipy \citep{jones2001scipy}, Matplotlib \citep{hunter2007matplotlib}, Astropy \citep{robitaille2013astropy,price2018astropy}, Healpy \citep{Zonca2019}, and Ray \citep{moritz2018ray}.

\bibliographystyle{unsrt}
\bibliography{main}
\appendix
\section{Theoretical signal and its demonstration on simulated maps}\label{sims-thoery}
In this section, we present the expected theoretical cross-correlation signal and how it matches our simulations. By comparing these two, we aim to demonstrate the validity of our cross-angular power spectrum approach in identifying anisotropic SGWB.  We first outline the theoretical estimate of the cross-correlation signal. Following this, we discuss the cross-correlation signal generated from our simulations. By comparing the theoretical predictions with the simulated signal, we can evaluate the consistency and robustness of our approach.

\subsection{Simulation-based estimate}

To simulate the anisotropic SGWB we employ the adaptive technique developed in \cite{sah2024imprints}. We use the results from different cosmological simulations as well as the analytical model of the binary evolution in the galactic environment to generate the population of SMBHBs. We use the SMBH mass ($M_{\rm BH}$) and Stellar mass ($M_{*}$) relation from the \texttt{ROMULUS} simulation \citep{saeedzadeh2023shining}. \texttt{ROMULUS} simulations are high-resolution, cosmological simulations, extending over a comoving volume of 25 $\rm Mpc^3$, that track the evolution of galaxies and SMBHs down to gravitational softening length of 0.7 kpc. We then use the $M_{\rm BH}$-$M_{*}$ relation to populate the \texttt{MICECAT}  \citep{crocce2015mice} simulated galaxy catalog with the SMBHBs using equations \eqref{M1}, \eqref{Mmu} and \eqref{q}. \texttt{MICECAT} is a halo and galaxy catalog derived from the large-scale cosmological simulation called \texttt{MICE-GC}, covering around 4 $\rm Gpc^3$ volume. 

We generate around $10^3$ realizations of the SMBHB population. Now, to obtain the total number of SMBHBs, emitting in the PTA band, we normalize the SGWB $\Omega_{\rm gw}(f)$ generated by the sources such that the mode of $\Omega_{\rm gw}(f)$ (over $10^{3}$ realizations) agrees with the power law curve \Big($\Omega_{\rm gw}(f) = B \times (f/f_{\rm yr})^{\alpha}$\Big) fitted to the 15-year data release of NANOGrav with the median value of the fitted parameters $B$ and $\alpha$. We generate simulated maps of the SGWB  by calculating the GW energy flux emitted by each binary of the simulated SMBHB population and then projecting the three-dimensional positions of galaxies and GW signal onto the 2d sphere.

\begin{figure}
    \centering
    \includegraphics[width=8cm]{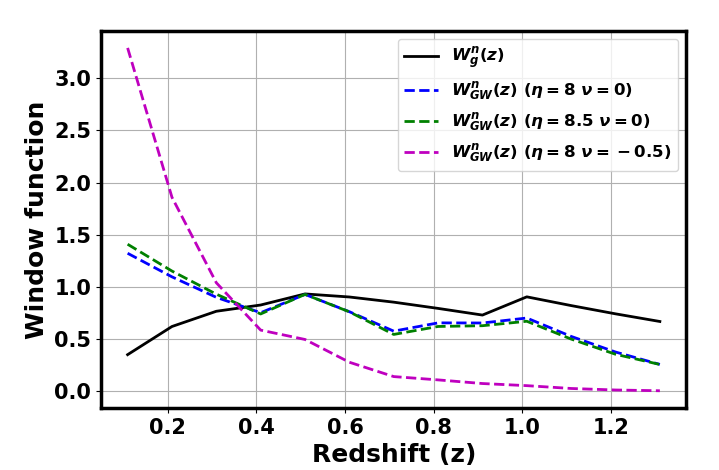}
    \caption{Normalised window function for \texttt{MICECAT} simulation, for galaxies, $W^{\rm n}_{\rm g}$, represented by solid line, and window function for SGWB, $W^{\rm n}_{\rm GW}$, represented by dash lines for different $M_{*}-M_{\rm BH}$ relation with fixed $\rho = 1$.}
    \label{Wn1}
\end{figure}

In Fig. \ref{Wn1}, we show the normalised window function of the galaxy ($W^{\rm n}_{\rm g}$) as well as the SGWB ($W^{\rm n}_{\rm GW}$). The window functions offer insights into the distribution and relative contributions of these sources across different redshifts. The window function of the galaxy peaks around z = 0.5 and then gently declines beyond this redshift. This pattern arises because the \texttt{MICECAT} catalog used here is complete for DES-like surveys ($i<24$) up to z=1.4. In contrast, the SGWB window function peaks near z=0 and continuously decreases with redshift. This trend occurs because low-redshift sources are brighter due to their closer distances compared to high-redshift sources.

\subsection{Theoretical estimation}
The theoretical cross-correlation signal is estimated using the Eq. \eqref{ClgGW_1}. The $\phi_{\rm g}(z)$ in the galaxy window function, $W_{\rm g}(z)$,  is calculated by taking the redshift distribution $\frac{dn}{dz}$ of the galaxy of the \texttt{MICECAT} catalog. Similarly, the  $\phi_{\rm GW}(z)$ in $W_{\rm GW}(z)$  is calculated by using Eq. \eqref{phiGW}, where the SMBHB mass distribution is obtained from the galaxy mass function of the catalog by assuming relations given by equations \eqref{M1}, \eqref{Mmu} and \eqref{q}. These window functions are then incorporated into the cross-correlation equation to estimate the theoretical $C_{\ell}^{\rm gGW}$.  We use the CAMB  python package \citep{lewis2011camb} to compute the matter power spectrum. CAMB allows us to calculate P(k) with high precision. By integrating the window functions over redshift spanned by the catalog, with the matter power spectrum at that redshift, we estimate the angular cross-correlation power spectrum, $C_{\ell}^{\rm gGW}$ capturing the contributions from different scales and redshift bins.

\begin{figure}
    \centering
    \includegraphics[width=0.45\textwidth]{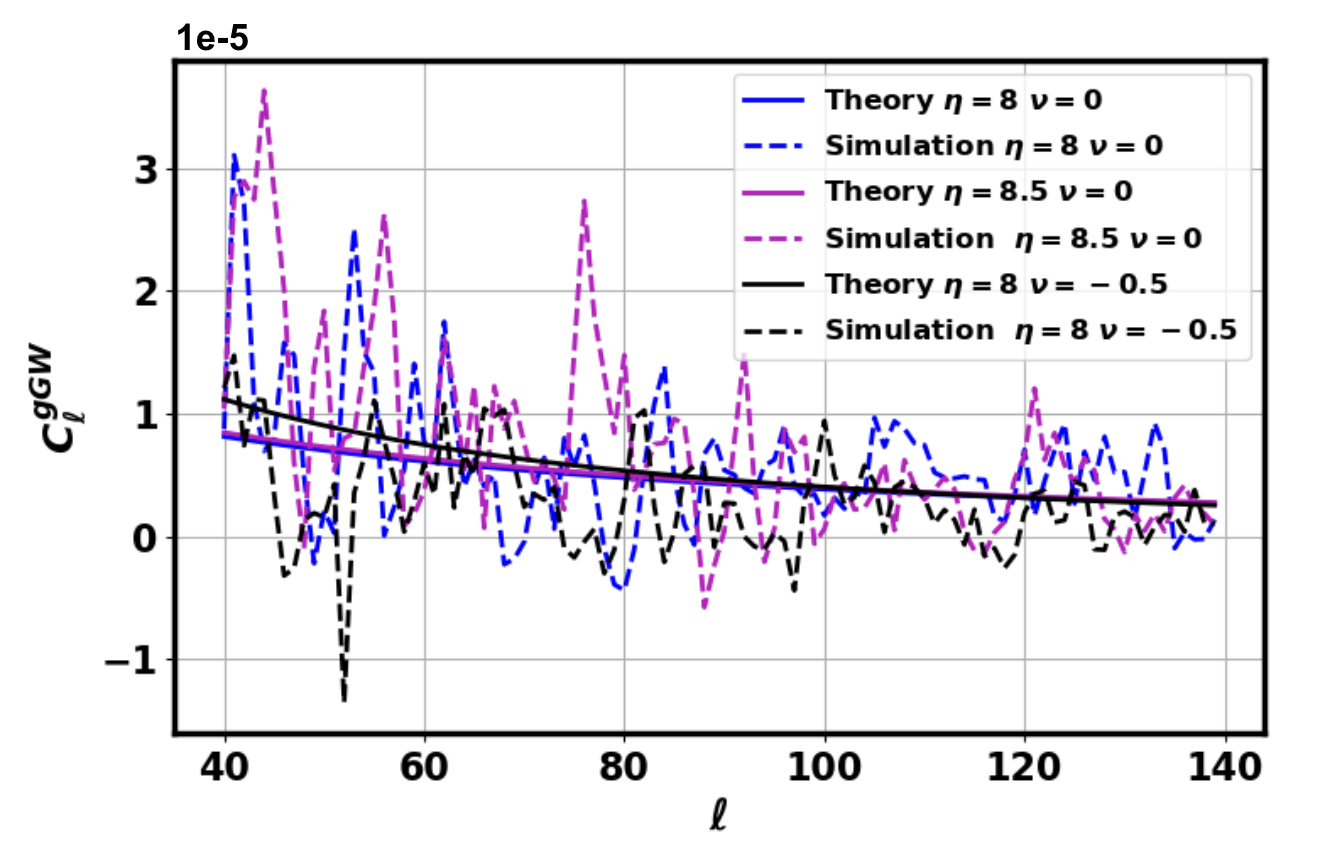}
    \caption{Theoretical $C_{\ell}^{\rm gGW}$ and Simulated $C_{\ell}^{\rm gGW}$ from MICECAT catalog for different $M_{*}-M_{\rm BH}$ relation with fixed $\rho = 1$. The simulated curves agree with the simulated curves.}
    \label{ClgGW_M}
\end{figure}

\subsection{Comparison of simulation-based estimation with theoretical estimation}
In Fig. \ref{ClgGW_M}, we illustrate the theoretical angular cross-correlation power spectrum, $C_{\ell}^{\rm gGW}$, as well as the simulated $C_{\ell}^{\rm gGW}$ obtained by taking the median over $10^{3}$ realizations of the source population. The plot demonstrates good agreement between the theoretical predictions and the simulated results, validating the cross-correlation formalism and the theoretical framework presented in previous sections. We have only demonstrated the curves for $\ell > 40$ because \texttt{MICECAT} covers only one-eighth of the sky, making it unreliable for small $\ell$ modes.

\section{Derivation of the angular noise power spectrum}\label{noise-estimator}
The overlap reduction function of the timing residual between two pulsars i and j is given by\citep{anholm2009optimal,Chamberlin:2014ria,Gair:2014rwa,Cornish:2014rva, pol2022forecasting}
\begin{equation}
    \begin{aligned}
        \Gamma_{ij} = & ~ \frac{3}{2} ~\sum_{k}  ~ P_{k} \frac{\Delta \omega_k}{4\pi}~ \Big[\sum_A \mathcal{F}^{A}_{i,k}  \mathcal{F}^{A}_{j,k}\Big],
    \end{aligned}
    \label{Gamma}
\end{equation}
where $\Delta\omega_{k}$ represents the sky area extended by pixel k, $ \mathcal{F}^{A}_{i}(\hat{n})$ and $ \mathcal{F}^{A}_{j}(\hat{n})$ are the response functions of the pulsars i and j, located in the directions $\hat{p}_i$ and $\hat{p}_j$, respectively, for gravitational waves coming from the direction $\hat{n}_{k}$. This is given by \citep{anholm2009optimal,siemens2013stochastic} 
\begin{equation}
     \mathcal{F}^{A}_{i,k} = \frac{1}{2} \frac{\hat{p_i}^{a}\hat{p_i}^{b}}{1+\hat{n_k}  \cdot \hat{p_i}} e^{A}_{ab},
\end{equation}

$e^{A}_{ab}$ is the gravitational wave polarization tensor, and $P_k$ represents the frequency-integrated power in SGWB at the sky pixel denoted by $k$, scaled by the mean SGWB. $P_k$ is defined as 
\begin{equation}
    P_k  \equiv \frac{\int \Omega^{k}_{\rm gw}(f) \frac{df}{f}}{\int \overline{\Omega}_{\rm gw}(f) \frac{df}{f}}, 
\end{equation}
where $\Omega^{k}_{\rm gw}(f)$ and $\overline{\Omega}_{\rm gw}(f)$ represent the SGWB density due to the sources in a unit solid angle at pixel k, and its mean, respectively.

We can represent Eq. \eqref{Gamma} in matrix form
\begin{equation}
        \bm{\Gamma} = \textbf{R} ~ \textbf{P},
\end{equation}

where $\textbf{R}^{ij}_{k}$  = $ \frac{3}{2} \times \frac{\Delta \omega_k}{4\pi}~ \Big[\sum\limits_A \mathcal{F}^{A}_{i,k}  \mathcal{F}^{A}_{j,k}\Big]  $ and $\textbf{P}_{k}$ = $P_{k}$.

The optimal timing cross-correlation estimator, the likelihood ratio maximized over the amplitude in the weak signal limit is given by \cite{anholm2009optimal,siemens2013stochastic,pol2022forecasting}
\begin{equation}
    \hat{\textbf{Y}}_{ij} = \frac{ \delta \textbf{t}_{i} \textbf{G}_{i}^{-1} \textbf{S}_{ij} \textbf{G}_{j}^{-1} \delta \textbf{t}_{j}}{\textbf{Tr}[\textbf{G}_{i}^{-1} \textbf{S}_{ij} \textbf{G}_{j}^{-1} \textbf{S}_{ij}]},
\end{equation}
where $\delta t_{i}$ represents the timing residual of the pulsar i, $\textbf{G}_{i}$ = $<\delta \textbf{t}_{i}$ $\delta \textbf{t}^T_{i}>$, and $\textbf{S}_{ij}$ $A^2$ $\Gamma_{ij}$ = $<\delta \textbf{t}^T_{i} \delta \textbf{t}^T_{j}>$, where A is the amplitude of the characteristic strain of the GW background. The characteristic strain $h_{c}(f) = A ~ (\frac{f}{f_{ref}})^{\gamma} $ is related to the $\Omega_{\rm gw}(f)$ as
\begin{equation}
    \Omega_{\rm gw}(f) = \frac{\pi}{4 G \rho_{\rm c}} f^{2} A^{2} \Bigg(\frac{f}{f_{\rm ref}}\Bigg)^{2\gamma},
\end{equation}
where $\rho_{\rm c}$ is the critical density of the universe,  $\gamma$ is the spectral index of the characteristic strain, and $f_{\rm ref}$ is the frequency at which the magnitude of the characteristic strain is A.

By defining an amplitude scaled estimator  $\hat{\textbf{X}}_{ij}$ = $\hat{\textbf{Y}}_{ij}$/$A^2$,  we can write the corresponding uncertainty on $\hat{\textbf{X}}_{ij}$ as
\begin{equation}
    \bm{\sigma}_{ij} = \frac{1}{A^2} (\textbf{Tr}[\textbf{G}_{i}^{-1} \textbf{S}_{ij} \textbf{G}_{j}^{-1} \textbf{S}_{ij}])^{-1/2}.
\end{equation}
In terms of $\hat{\textbf{X}}_{ij}$ , we can write the likelihood as \cite{pol2022forecasting} 
\begin{equation}
    \mathcal{P}(\hat{\textbf{X}}|\textbf{P}) \propto \textbf{exp}[\frac{-1}{2}(\hat{\textbf{X}} - \textbf{R} \textbf{P})^{T} \bm{\Sigma}^{-1} (\hat{\textbf{X}} - \textbf{R} \textbf{P})],
\end{equation}
where $\bm{\Sigma}$ is the timing cross-correlation covariance matrix. The minimum  variance estimator for the sky signal in the pixel ($ \hat{\textbf{P}}$) is \citep{thrane2009probing,pol2022forecasting} 
\begin{equation}
    \hat{\textbf{P}} = \mathbf{\Sigma}_{\rm pix}\textbf{R}^{T}\mathbf{\Sigma}^{-1} \hat{\textbf{X}},
\end{equation}
where, the covariance matrix on $\hat{\textbf{P}}$ is
\begin{equation}
    \begin{aligned}
        \mathbf{\Sigma}_{\rm pix} = &~  \textbf{R}^{-1} \bm{\Sigma} \textbf{R}.
    \end{aligned}
\end{equation}
In the spherical harmonic basis, we can write $\hat p_{lm}= \int d^2\hat n \text{Y}^*_{lm}(\hat n) \text{P}(\hat n)$. The corresponding noise angular power spectrum in the spherical harmonic basis is 
\begin{equation}
    \begin{aligned}
        \textbf{N}_{\ell m\ell'm'}  =\textbf{R}_{(\ell m)}^{-1} \bm{\Sigma} \textbf{R}_{(\ell' m')}. 
    \end{aligned}\label{noise1}
\end{equation}
This quantity depends on the pulsar-timing noise covariance matrix $\Sigma$ and the response function in harmonic space denoted by $\textbf{R}_{(\ell m)}$ which depends on the number of pulsars and their spatial distributions. For an isotropic distribution of pulsars with same pulsar timing residual noise for every pulsars, the above quantity becomes diagonal and can be written as \citep{Cornish:2014rva,Roebber:2016jzl}
\begin{equation}
    \begin{aligned}
        \textbf{N}_{\ell m\ell'm'}  \delta_{\ell\ell'}\delta_{mm'}\equiv  N_{\ell}  =& \frac{4\pi \sigma^2_{\rm pix} }{N_{\rm pix}}e^{\frac{\ell (\ell+1) \theta^2_{\rm b}}{8\rm ln2}},
    \end{aligned}
\end{equation}
where $N_{\rm pix}$ represents the number of pixel which satisfies the condition $N_{\rm pix} \leq$ N$_{\rm pairs}$ and $\sigma_{\rm pix}$ is standard deviation of pixel noise. The quantity in the exponential comes from the maximum number of modes on the sky that are accessible from a given set of pulsars. The maximum accessible mode denoted by $\ell_{\rm max}$ is limited by the number of pulsars $N_{\rm p}$ by $\leq N_{\rm p}^{1/2}$ (equal for homogeneous distribution of pulsars)\citep{pol2022forecasting}. For $\ell$-modes larger than $\ell_{\rm max}$, we model the noise using a Gaussian beam with a full-width half maxima given by $\theta_b = \pi/\ell_{\rm max}$. For an inhomogeneous distribution of pulsars, the noise covariance matrix becomes sky-position dependent and the beam function will become non-Gaussian. For an exact distribution of the pulsars from SKA \cite{smits2009pulsar}, we can estimate the complete noise covariance matrix $\textbf{N}_{\ell m\ell'm'}$ without any assumption using Eq. \eqref{noise1}. For such scenarios, the effective number of sky pixels will be less than N$_{\rm pairs}$, and the noise will be large. Also, the number of independent $\ell$ modes will not be $N_p^{1/2}$. In order to capture the scenarios of inhomogeneous distribution of pulsars and unequal noise of different pulsars (due to their intrinsic red-noise \citep{alam2020nanograv,johnson2024nanograv}, we model the noise in terms of two parameters $\sigma$ and $\ell_{\rm max} = \pi/\theta_b$ as
\begin{equation}
N_{\ell}  = \sigma^2 e^{\frac{\ell (\ell+1) \theta^2_{\rm b}}{8\rm ln2}}. 
\end{equation}
The parameter $\sigma^2$ (in units of steradian) is an effective parameter for noise which can be a large value even when there are a large number of pulsars, allowing for a scenario of inhomogeneous distribution of pulsars and their unequal timing residual noise $\sigma_{\rm pix}$ of different pulsars. The parameter $\ell_{\rm max}$ captures the effective number of independent modes that can be measured on the sky, and need not be equal to $N_{\rm p}^{1/2}$, allowing for an inhomogeneous distribution. In this analysis, we vary ($\sigma, \ell_{\rm max}$) from $(1, 5)$ (for current PTA \cite{pol2022forecasting}) to ($10^{-3}, 80$) (accessible in future from SKA \cite{smits2009pulsar}).

\end{document}